\documentclass{article}
\usepackage{spconf,graphicx,amssymb,amsmath,epsfig,algorithm,algorithmic,color,soul}
\def\1{\mathbf{1}}
\def\a{\mathbf{a}}

\def\b{\boldsymbol{\beta}}

\def\g{\boldsymbol{\gamma}}

\def\N{\mathcal{N}}

\def\T{\mathsf{T}}

\def\x{\mathbf{x}}

\def\y{\mathbf{y}}

\def\z{\mathbf{z}}

\def\1{\textbf{1}}
\def\0{\textbf{0}}

\newcommand{\bi}{\begin{itemize}}
\newcommand{\ei}{\end{itemize}}
\newcommand{\be}{\begin{equation}}
\newcommand{\ee}{\end{equation}}
\newcommand{\ben}{\begin{equation*}}
\newcommand{\een}{\end{equation*}}
\newcommand{\bcs}{\begin{cases}}
\newcommand{\ecs}{\end{cases}}
\newcommand{\bpm}{\begin{pmatrix}}
\newcommand{\epm}{\end{pmatrix}}
\newcommand{\bmm}{\begin{matrix}}
\newcommand{\emm}{\end{matrix}}
\newcommand{\bc}{\begin{center}}
\newcommand{\ec}{\end{center}}

\DeclareGraphicsExtensions{.eps}

\def\N{\mathcal N}
\def\l{\lambda}
\def\g{\gamma}
\def\a{\alpha}
\def\aa{\boldsymbol\alpha}
\def\b{\beta}
\def\st{\sigma_t}
\def\sn{\sigma_n}
\def\sa{\sigma_a}
\def\sps{\sigma_{ps}}
\def\H{\mathcal H}
\def\jh{\hat\jmath}
\newcommand{\argmin}[1]{\underset{#1}{\operatorname{arg\,min\:}}}

\providecommand{\abs}[1]{\lvert#1\rvert}
\providecommand{\norm}[1]{\lVert#1\rVert}
\def\m{\boldsymbol{m}}
\def\T{\mathsf{T}}
\def\v{\boldsymbol{v}}
\def\y{\boldsymbol{y}}
\def\z{\boldsymbol{z}}

\title{
	Location-aided Distributed Primary User Identification\\
	in a Cognitive Radio Scenario
}
\name{
	Pavle Belanovic, Sergio Valcarcel Macua, and Santiago Zazo
	\thanks{
		This work was supported in part by the Spanish Ministry of Science and Innovation under the grant TEC2009-14219-C03-01; the Spanish Ministry of Science and Innovation in the program CONSOLIDER-INGENIO 2010 under the grant CSD2008-00010 COMONSENS; the European Commission under the grant FP7-ICT-2009-4-248894-WHERE-2; the European Commission under the grant FP7-ICT-223994-N4C and the Spanish Ministry of Science and Innovation under the complementary action grant TEC 2008-04644-E; Spanish Ministry of Science and Innovation under the grant TEC2010-21217-C02-02-CR4HFDVL.
	}
}
\address{
	Escuela T\'ecnica Superior de Ingenieros de Telecomunicaci\'on\\
	Universidad Polit\'ecnica de Madrid
}
\begin{document}

\ninept
\addtolength{\itemsep}{-1mm}
\addtolength{\abovedisplayskip}{-3pt}
\addtolength{\belowdisplayskip}{-3pt}
\addtolength{\abovecaptionskip}{-5pt}
\addtolength{\belowcaptionskip}{-12pt}

\newcommand{\ssection}[1]{%
  \vspace{-3pt}\section[#1]{#1}\vspace{-3pt}}

\maketitle
\begin{abstract}
	We address a cognitive radio scenario, where a number of secondary users performs identification of which primary user, if any, is transmitting, in a distributed way and using limited location information. We propose two fully distributed algorithms: the first is a direct identification scheme, and in the other a distributed sub-optimal detection based on a simplified Neyman-Pearson energy detector precedes the identification scheme. Both algorithms are studied analytically in a realistic transmission scenario, and the advantage obtained by detection pre-processing is also verified via simulation. Finally, we give details of their fully distributed implementation via consensus averaging algorithms.
\end{abstract}
\ssection{Introduction}
\label{sec:intro}
In modern wireless networks, radio spectrum is a precious resource. Cognitive radio is one method of making ad-hoc use of unoccupied spectrum in order to increase the efficiency of its use. At the core of this approach lies the problem of detecting, and identifying, active primary users by a network of secondary users. We study the identification of which, if any, primary user is transmitting, by a network of secondary nodes without a fusion center and with only elementary location information. 
In a network of decision makers, distributed detection has been thoroughly studied and different solutions have been proposed. The problem is to decide what information the agents should share, and to find optimal fusion rules to combine the local outputs. Decentralized binary detection~\cite{Tsitsiklis1993,Varshney1997,Blum1997,Viswanathan1997,Chamberland2003,Quan2010} proposes a parallel architecture in which every node sends a summary of its own observations (e.g. quantized values, test outputs or hard decisions) to a fusion center in charge of making the final decision. Recently, completely distributed implementations, in which there is no fusion center so the nodes have to collaborate with each other to converge to the global solution~\cite{Kar2007}, have also appeared; paying special attention to on-line algorithms in which nodes collaborate and detect in the same timescale~\cite{Braca2010,Cattivelli2011,Bajovic2011}. 

The M-ary hypothesis testing, in particular with no prior knowledge of the probability distributions of the alternative hypothesis, has received much less attention. A number of decentralized approaches, which rely on a fusion center, have been proposed. For instance,~\cite{Liu2011} applies a blind algorithm after estimating the prior probabilities of the hypothesis; while in~\cite{Zhang2001} the M-ary detection problem is converted into a sequence of binary detection problems. A fully distributed scheme based on belief propagation has been proposed in~\cite{Saligrama2006}, but it requires knowledge of the prior probabilities in order to maximize the posterior distribution.

In this paper we introduce two fully distributed algorithms for transmission detection and primary user identification (M-ary hypothesis) when the only prior knowledge is that of the noise distribution. Nevertheless, we make the assumption that rudimentary location information is available: each secondary node knows its attenuation factor from each primary user. This assumption is reasonable in practical scenarios, because the nodes can easily learn the attenuations though calibration (indoor or outdoor, static only), fingerprinting (indoor, static or dynamic), GPS location and a propagation model (outdoor, static or dynamic), or any other method, all of which are beyond the scope of this paper.
\ssection{Problem definition}
\label{sec:problem}
Let us assume a cognitive radio scenario, where $P$ primary users and $S$ secondary users share the same geographic area. Each primary user may transmit at any time (though we assume at most one primary user is transmitting at any moment) using a random "bursty" transmission. Each such transmission by a primary node $p$ is modeled as a signal $s_p$ which alternates between an \emph{active} and a \emph{passive} state, whose lengths are Poisson random variables, with parameters $\l q$ and $(1-\l)q$, such that $\lambda$ is the activity factor and $q$ is the expected number of samples in each cycle. During the active state, $s_p\sim\N(0,\st^2)$, and in the passive state $s_p=0$. For each transmission, the primary node selects random $\st^2$ and $\l$, unknown to the secondary nodes.

The transmitted signal is then propagated to all the secondary users. Here we are not concerned with any propagation model in particular. Instead, we assume a static model of the received signal $x_s$ at the secondary node $s$ as an attenuation of the transmitted signal $s_p$ in AWGN, \hbox{$x_s=\a_{ps}s_p+n$}, where \hbox{$n\sim\N(0,\sn^2)$}. We assume the realizations of $n$ are iid at all the secondary nodes, and each node estimates $\sn^2$ perfectly. Globally, the attenuation is given by a matrix \hbox{$A=[\a_{ps}]_{P\times S}$}, where each coefficient is assumed static and derived by any means, e.g. geometric model, measurements, fingerprinting technique, etc. We assume that each node $s$ has complete knowledge of its column of $A$, but not of any other nodes' attenuations. Other than this, no further location information is required by any node, neither its own location, nor of any other (primary or secondary) node. 

Given this model, we tackle the problem of identifying the transmitting primary user, if any, by means of a distributed algorithm that does not rely on the availability of a fusion center serving the network of secondary nodes. In other words, the nodes must cooperatively decide among \hbox{$P+1$} hypotheses \hbox{$\{\H_0,\H_1,\ldots,\H_P\}$}, where \hbox{$\H_0$} represents no transmission from any primary. To this end, we propose two suitable algorithms, presented in the following sections.
\ssection{Identification scheme}
\label{sec:identify}
The first scheme we propose performs direct identification based on distributed hypothesis testing. Each node $s$ in isolation performs energy sampling, where $W$ integration windows, each of length $L$ samples, produces an energy estimate \hbox{$y_s[w]=\frac{1}{L}\sum_{l=1}^L(x_s[l])^2$}, with \hbox{$w\in\{1,2,\ldots,W\}$}. Using the knowledge of the noise statistic, we generate a new variable \hbox{$\z_s=\y_s-\sn^2$}, distributed
\footnote{
	Throughout we approximate $\chi^2$ distributions with $L$ (and later $W$) degrees of freedom by Gaussian distributions with the same first two moments.
}
as
\ben
	\z_s\sim
	\bcs
		\N\left(0,\frac{2\sn^4}{L}\right) & \H_0\\
		\N\left(\sps^2\l,\frac{2}{L}(\sn^4+\sps^4\l+2\sn^2\sps^2\l)\right) & \H_p
	\ecs
\een
where \hbox{$\sps=\a_{ps}\st$} and \hbox{$p\in\{1,2,\dotsc,P\}$}. Each node $s$ then constructs \hbox{$P+1$} hypotheses to test, by compensating its own received distribution of $W$ samples of $\z_s$ exactly \hbox{$P+1$} times. The first compensation represents $\H_0$, i.e. the possibility that $\z_s$ contains only noise energy, and is constructed simply by using the raw data itself (no compensation). The following $P$ compensations are performed by multiplying the received distribution by a compensation factor \hbox{$\b_p\a_{ps}^{-2}$}, i.e. one compensated distribution for each possible primary node. The factor $\b_p$ serves to normalize each of the hypothesis, relative to $\H_0$, so that later on their variances will be directly comparable. Hence, \hbox{$\b_p=\left(\norm{\aa_p}\right)^{-1/2}$}, where \hbox{$\aa_p=[\a_{p1},\a_{p2},\dotsc,\a_{pS}]^\T$}.

Therefore, assuming that the hypothesis $\H_p$ is true (shown in bold), each node $s$ has a set of compensated distributions
\begin{align}
	\N\left(\sps^2\l,\frac{2}{L}\left(\sn^4+\sps^4\l+2\sn^2\sps^2\l\right)\right) \quad& \H_0\notag\\[-1pt]
	\N\left(\frac{\b_1\sps^2\l}{\a_{1s}^2},\frac{2\b_1^2}{\a_{1s}^4L}\left(\sn^4+\sps^4\l+2\sn^2\sps^2\l\right)\right) \quad& \H_1\notag\\[-1pt]
	& \dotsm\notag\\[-7pt]
	\boldsymbol{\N\left(\b_p\st^2\l,\frac{2\b_p^2}{\a_{ps}^4L}\left(\sn^4+\sps^4\l+2\sn^2\sps^2\l\right)\right)} \quad& \boldsymbol{\H_p}\notag\\[-1pt]
	& \dotsm\notag\\[-7pt]
	\N\left(\frac{\b_P\sps^2\l}{\a_{Ps}^2},\frac{2\b_P^2}{\a_{Ps}^4L}\left(\sn^4+\sps^4\l+2\sn^2\sps^2\l\right)\right) \quad& \H_P\notag
\end{align}
Hence, there are $S$ distributions for each hypothesis, one per node.

Another way of seeing this is that each node estimates, in isolation, the product of the only two parameters common to all, \hbox{$\st^2\l$}, since \hbox{$\text{E}(\z_s|_{\H_p}/\a_{ps}^2)=\st^2\l\;\forall s$}, when primary $p$ is transmitting. We note that for the correct hypothesis $p$, though all the nodes agree perfectly in the mean, they do not in the variance, due to the different attenuations $\a_{ps}$ which do not disappear.

Estimating which hypothesis is true, $\jh$, across the $S$ secondary nodes is the next challenge, and the first to use coordination among the secondary nodes. An intuitive approach would be to choose the hypothesis with a minimum sum of distances among the $S$ distributions. Remembering that in the correct hypothesis, the variance in all the nodes does not match, prevents us from using Bhattacharyya, Mahalanobis, or any distance metrics that take variance into consideration. In other words, we disregard the variance information in each node, and use only on the means. Hence, we opt for the Euclidean metric, such that
\begin{align}
	\jh&=\argmin{j}\sum_{m=1}^S\sum_{n=1}^S\;\abs{\mu_{m|_{\H_j}}-\mu_{n|_{\H_j}}}\notag\\[-2pt]
	&=\argmin{j}\;\b_p\st^2\l\sum_{m=1}^S\sum_{n=1}^S\left|\left(\frac{\a_{pm}}{\a_{jm}}\right)^2-\left(\frac{\a_{pn}}{\a_{jn}}\right)^2\right|\notag
\end{align}
It is of course easy to see that the function reaches its minimum, being $0$, when \hbox{$j=p$} and is strictly greater otherwise. Unfortunately this problem formulation cannot be used in a distributed scenario because it requires the knowledge of the entire matrix $A$.

\begin{figure*}
	\centering
	\includegraphics[width=.77\textwidth]{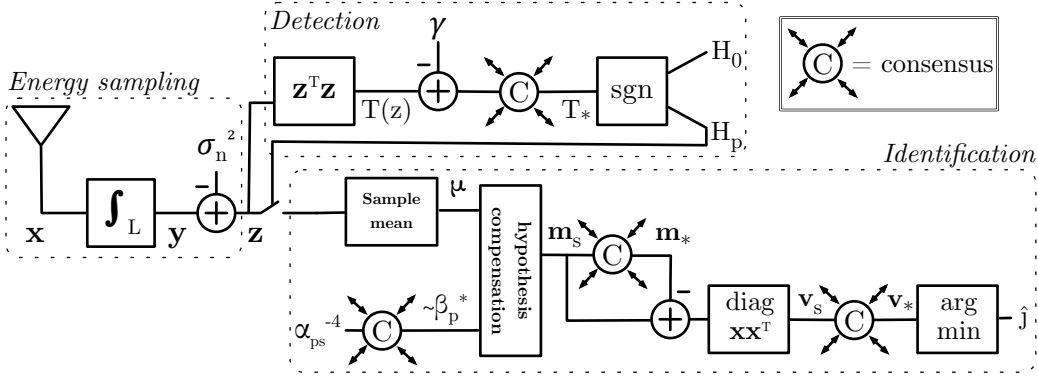}
	\caption{Structure of the identification scheme with detection pre-processing}
	\label{fig:structure}
\end{figure*}

It is also easy to show that this sum of distances is proportional to the sample variance of the set of $S$ compensated means. Hence, the problem reduces to finding the hypothesis with minimum variance across the $S$ nodes. This problem is easily tackled in a distributed fashion using averaging consensus algorithms~\cite{Garin11a}, following the idea of constructing the sample covariance matrix shown in~\cite{Valcarcel11a}. Hence, our proposed algorithm is shown in Algorithm~\ref{alg:ident}, and illustrated in Fig.~\ref{fig:structure}, where the detection block is \emph{not} active.

\begin{algorithm}
	\caption{Identification algorithm at node $s$}
	\label{alg:ident}
	\begin{algorithmic}[1]
		\setlength{\itemsep}{.25em}
		\STATE \textbf{INPUT} $\;\x_s,\,\a_{ps}\forall p,\,\sn,\,L,\,W$
		\STATE $y_s[w]\gets L^{-1}\x_s^\T\x_s$ (for each of the $W$ elements of $\y_s$) \label{alg:energystart}
		\STATE $\z_s\gets\y_s-\sn^2$
		\STATE $\mu_s\gets W^{-1}\z_s^\T\1$ (sample mean of $\z_s$)
		\STATE $\b_p^*\gets \frac{1}{S}\sum_{i=1}^S \a_{pi}^{-4}$\hfill$\Longleftarrow\;$\emph{consensus loop} \label{alg:energyend}
		\STATE $\m_s\gets [\mu_s,\mu_s\b_1\a_{1s}^{-2},\dotsc,\mu_s\b_P\a_{Ps}^{-2}]^\T$ \label{alg:means}
		\STATE $\m_*\gets \frac{1}{S}\sum_{i=1}^S \m_i$\hfill$\Longleftarrow\;$\emph{consensus loop} \label{alg:identstart}
		\STATE $\v_s\gets \text{diag}((\m_s-\m_*)\times(\m_s-\m_*)^\T)$
		\STATE $\v_*\gets \frac{1}{S}\sum_{i=1}^S \v_i$\hfill$\Longleftarrow\;$\emph{consensus loop} 
		\STATE $\jh\gets\argmin{j}\;v_j\in\v_*$ (index of the minimum element) \label{alg:identend}
		\STATE \textbf{OUTPUT} $\;\jh$
	\end{algorithmic}
\end{algorithm}

\ssection{Detection pre-processing}
\label{sec:detection}
In a low SNR regime the mode of failure of the scheme presented in Section~\ref{sec:identify}, and indeed any identification scheme, is that of always choosing $\H_0$, even when a signal is present, simply because this signal is too weak to identify a particular primary node transmitting. Hence, it makes sense to perform an (optimal or nearly-optimal) detection step first, detecting the activity of \emph{any} primary node, followed then by an identification procedure similar to that of Section~\ref{sec:identify}, but this time with only $P$, rather than \hbox{$P+1$} hypotheses. This is illustrated in Fig.~\ref{fig:structure}.

If each node $s$ had the knowledge of all the parameters of its $\z_s$, the optimal test~\cite{Kay98a} based on the Neyman-Pearson criterion would be given by \hbox{$T(\z_s)=\eta\z_s^\T\1+\theta\z_s^\T\z_s$}. Both $\eta$ and $\theta$ are functions of the statistics of $\z_s$, and since these are not known, the approach is not feasible. It is in principle possible to construct the generalized likelihood ratio test (GLRT) using the maximum likelihood (ML) estimates of $\sps^2$ and $\l$. However, we found that in this cognitive radio scenario the GLRT performed poorly, since the estimates were very poor in the hypothesis $\H_0$.

Therefore, we propose a sub-optimal approach $T(\z_s)=\z_s^\T\z_s$
\ben
	T(\z_s)\sim
	\bcs
		\N\left(\frac{2W\sn^4}{L},\frac{8W\sn^8}{L^2}\right) & \H_0\\
		\N\left(W(\sa^2\!+\!\sps^4\l^2),2W\sa^2(\sa^2\!+\!2\sps^4\l^2)\right) & \H_p
	\ecs
\een
where $\sa^2=\frac{2}{L}(\sn^4+\sps^4\l+2\sn^2\sps^2\l)$ for compactness. One obvious advantage is that this test does not depend on the estimates of $\sps^2$ and $\l$.

The threshold is hence \hbox{$\g=\frac{2\sn^4}{L}(\sqrt{2W}Q^{-1}(P_{fa})+W)$}, and is calculated by each node in isolation, for a defined probability of false alarm $P_{fa}$, where $Q^{-1}(\cdot)$ is the inverse Q-function. Typically each node would compare the local \hbox{$T(\z_s)\lessgtr\g$} producing a 1-bit detection decision, which are then combined globally (e.g. voting~\cite{Aalo1992}).

Instead, we propose a \emph{weighted} global test \linebreak\hbox{$T_*=\frac{1}{S}\sum_{i=1}^S(T(\z_s)-\g)\lessgtr0$}. Although the factor $S^{-1}$ is quite unnecessary, it shows that this global value can also be derived in a distributed fashion via average consensus. If each node calculates its \emph{vote} as a degree of confidence \hbox{$T(\z_s)-\g$}, simply the sign ($+$ or $-$) of the global average of the votes (available at all the nodes simultaneously) is the outcome of the global test. This weighing allows the nodes closer to any transmitting primary to exert a bigger influence, as desired.

Once the detection stage is performed in this distributed fashion, all the nodes can carry out the identification procedure (also distributed) as shown in Section~\ref{sec:identify}, but this time with $P$ rather than \hbox{$P+1$} hypotheses. This is shown in Algorithm~\ref{alg:detect} and illustrated in Fig.\ref{fig:structure}, where the detection block is active.
\begin{algorithm}
	\caption{Identification algorithm with pre-detection at node $s$}
	\label{alg:detect}
	\begin{algorithmic}[1]
		\setlength{\itemsep}{.25em}
		\STATE \textbf{INPUT} $\;\x_s,\,\a_{ps}\forall p,\,\sn,\,L,\,W,\,P_{fa}$
		\\\vspace{\itemsep} \qquad \emph{--- Lines~\ref{alg:energystart} to~\ref{alg:energyend} of Algorithm~\ref{alg:ident} ---}
		\STATE $\g\gets\frac{2\sn^4}{L}(\sqrt{2W}Q^{-1}(P_{fa})+W)$ (NP threshold, local calc.)
		\STATE $T(\z_s)\gets\z_s^\T\z_s$ (simplified local test)
		\STATE $T_*\gets\frac{1}{S}\sum_{i=1}^S(T(\z_s)-\g)$\hfill$\Longleftarrow\;$\emph{consensus loop} 
		\IF{$T_*<0$}
			\STATE $\jh=0$ (no transmission, $\H_0$)
		\ELSE
			\STATE $\m_s\gets [\mu_s\b_1\a_{1s}^{-2},\dotsc,\mu_s\b_P\a_{Ps}^{-2}]^\T$
			\\\vspace{\itemsep} \qquad \emph{--- Lines~\ref{alg:identstart} to~\ref{alg:identend} of Algorithm~\ref{alg:ident} ---}
		\ENDIF
		\STATE \textbf{OUTPUT} $\;\jh$
	\end{algorithmic}
\end{algorithm}
\ssection{Experiments}
\label{sec:experiments}
We verify, illustrate, and compare the functioning of the two proposed approaches via simulations.\footnote{Following the idea of reproducible research, the Matlab code for these experiments will be made available for download with the final article.} We define a simple scenario with four primary $(P=4)$ and twenty secondary users $(S=20)$ uniformly randomly located in a square area with sides of $200\,m$, choosing the primary users to be the most distant nodes. As discussed earlier, we assume iid zero-mean Gaussian noise at every secondary node, with the variance $\sn^2$ perfectly estimated by the secondary users. Since every node has a different signal to noise ratio (SNR), depending on its attenuation $\a_{ps}$, it is not possible to express the global results against SNR. Instead, we analyze the influence of $\st$ (keeping $\sn$ constant), or quite equivalently the transmitted-signal to receiver-noise ratio $S_tN_rR=20\log(\st/\sn)$, which is a fictitious parameter. We ran $10^4$ experiments for each value of \hbox{$S_tN_rR$} in the range from $0$ to $80\,\text{dB}$. In each experiment, we choose one of the \hbox{$P+1$} equally probable hypotheses. The identification scheme uses $100$ integration windows $(W=100)$ with two hundred samples each $(L=200)$. The transmitter activity factor is $50\%$ $(\lambda=0.5)$, with on average twenty samples per cycle $(q=20)$.

\begin{figure*}[htb]
\begin{minipage}[b]{\columnwidth}
  \centering
  \centerline{\includegraphics[width=.92\textwidth]{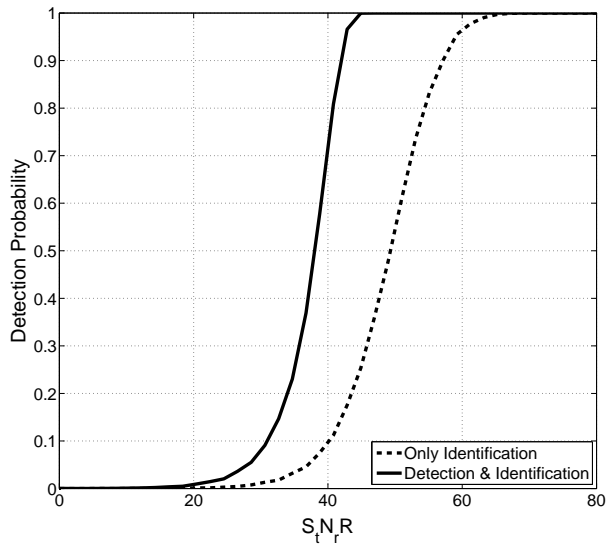}}
  \centerline{(a) Probability of detection}\medskip
\end{minipage}\hspace{3mm}
\begin{minipage}[b]{\columnwidth}
  \centering
  \centerline{\includegraphics[width=.92\textwidth]{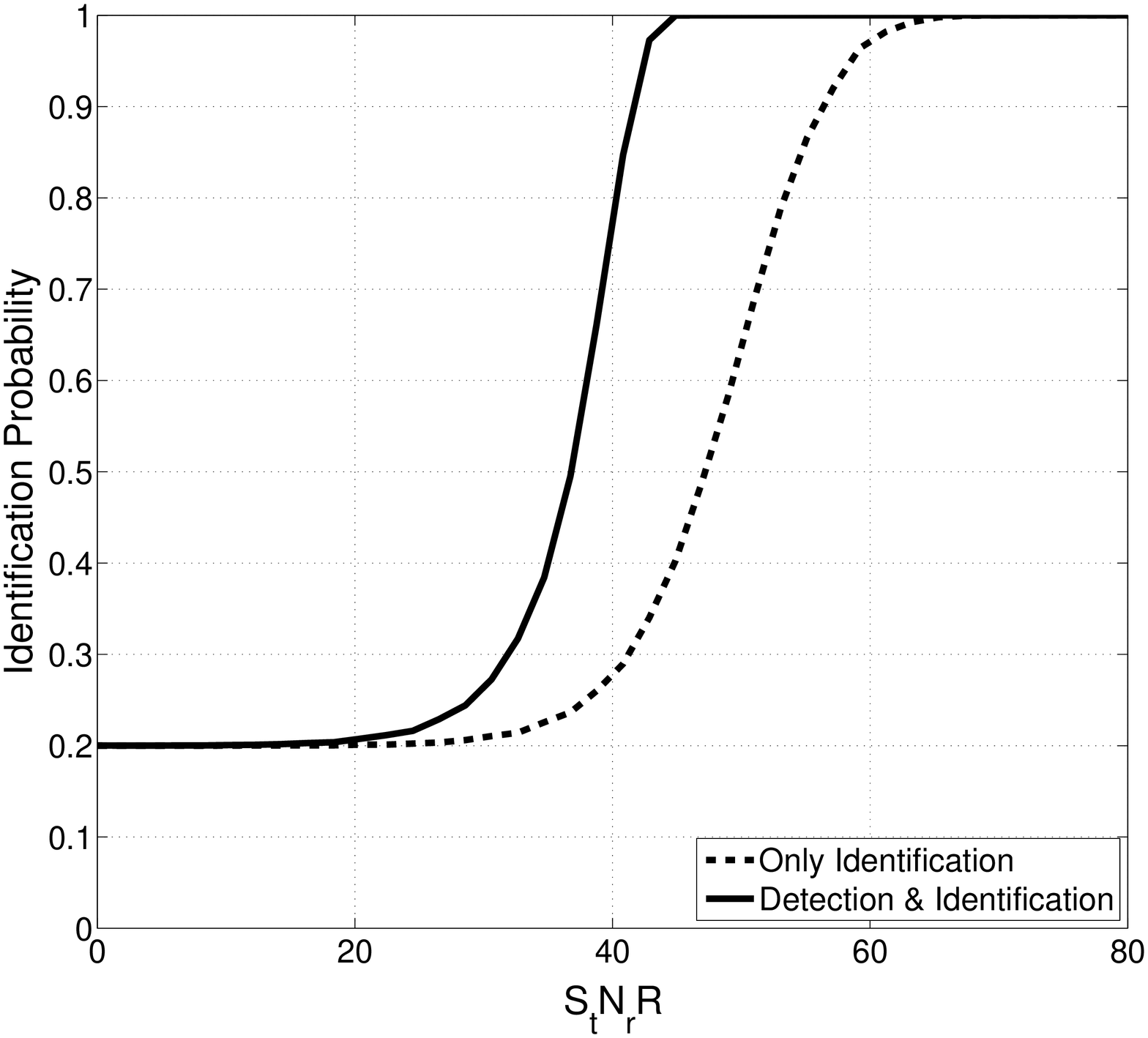}}
  \centerline{(b) Probability of identification}\medskip
\end{minipage}
\caption{Probability of detection and identification with increasing levels of $S_tN_rR$}
\label{fig:perf}
\end{figure*}

In Fig.~\ref{fig:perf}(a) we show the performance of both schemes in terms of a classic metric, the probability of detection. We see that at low to mid $S_tN_rR$ levels, the nearly-optimal detection pre-processing step brings about $10\,\text{dB}$ of improvement. Both curves converge on the left to the value of the $P_{fa}$, as usual, since both schemes fail in the same way. They are unable to separate the $\H_0$ and $\H_p$ hypotheses, which at such low $S_tN_rR$ levels practically overlap completely.

On the other hand, in Fig.~\ref{fig:perf}(b) we show the probability of identification, being the proportion of successful hypothesis identifications relative to the total number of experiments at that hypothesis, summed over all the hypotheses and normalized by \hbox{$(P+1)^{-1}$}. Again, we see a significant improvement won by the pre-detection step, of around $10\,\text{dB}$. As expected, both curves converge on the left to \hbox{$(P+1)^{-1}$}, which is \hbox{$1/5$} in this experiment.
\ssection{Conclusions}
\label{sec:conclusions}
In this work we studied the problem of identifying the active primary user by a distributed network of secondary nodes with very limited location information. We proposed two fully distributed algorithms, based on identification only, or identification with a detection pre-processing step. In the detection phase, we introduced a novel weighted global test, which allows the secondary nodes closer to the transmitter to exert a bigger influence. Both algorithms are implemented using averaging consensus to provide coordination among the nodes. As expected, the nearly optimal detection step brings a compelling improvement of about $10\,\text{dB}$. Future work on this topic may include constructing hypotheses for multiple simultaneously transmitting primary users, and exploring the effect of imperfect knowledge (estimates) of the attenuation factors.
\nocite{*}
\bibliographystyle{IEEEbib}
\bibliography{refs}
\end{document}